\documentclass[12pt,a4paper]{report}
\usepackage{graphics}
\begin{document}
\baselineskip 0.25in

\begin{center}
{\Large{\bf{Quasi-one dimensional graphite ribbon structures in the presence of
    a magnetic field and the on-site Coulomb correlation at half-filling}}}\\

Jayeeta Chowdhury$^1$, Shreekantha Sil$^2$, S. N. Karmakar$^3$, Bibhas
Bhattacharyya$^{4,5}$\\

\small{$^1$Department of Physics, Scottish Church College, 
1 \& 3 Urquhart Square, Kolkata 700 006, India}\\
\small{$^2$Department of Physics, Visva-Bharati, 
Santiniketan, West Bengal, 731235, India}\\
\small{$^3$TCMP Division, Saha Institute of Nuclear Physics, 
1/AF Bidhannagar, Kolkata 700 064, India}\\
\small{$^4$Department of Physics and Astrophysics, 
West Bengal State University, Barasat, Kolkata 700 126, India}\\
\small{$^5$Department of Physics, Jadavpur University, Kolkata 700 032, India}\\
\end{center}

{\bf Shortened version of the title:}\\
Graphite ribbons in presence of magnetic field and Hubbard correlation
\newpage
\begin{abstract}
We have presented the role of the Coulomb interaction ($U$) and the magnetic
field ($\vec{B}$) on the ground state properties of the quasi-one dimensional
graphite ribbon structures at half-filling. Mean field Hartree-Fock
Approximation is used to study the systems. To understand the boundary effects
in graphite structures, we have compared the results of these systems
with those of the square lattice ribbon structures. Studying the density of
states, the Drude weight and the charge gap, we have drawn the $U-B$ phase
diagrams for the zigzag and the armchair graphite ribbons. 
\end{abstract}

\section{Introduction}
Recently, there has been a lot of theoretical work on the carbon-network
structures due to the scope of wide applicability of the materials composed of
carbon atoms in nanotechnology devices \cite{Ijima}. Study
of nanoscale graphites seems interesting since in these systems edge and bulk
effects are comparable. States near the Fermi level have strikingly different
features depending on the boundary geometry of the graphites. The graphite
lattices with finite widths, known as graphite ribbons, are the simplest
systems to study the boundary effects. There are two basic edge shapes in
graphites, the armchair and the zigzag. Zigzag ribbons posses localized edge
states at the Fermi level, which correspond to the
non-bonding molecular orbital \cite{Nakada, Fujita, Waka1}. These types of
edge states are completely absent in armchair ribbons. It was observed, both
experimentally and theoretically, that depending on the boundary geometry and
size, graphite ribbons may be metallic or insulating 
\cite{Hamada, White, Mintmire, Wild, Odom, Nov1}. In the recent past, there
has been 
a lot of exciting experimental work on the graphenes exploring their response
to electric and magnetic fields \cite{Geim, Zhang}. 

A lot of work has been done studying the magnetic properties of the graphite
ribbons \cite{Lu, Ajiki, Lin, Bach}. It was observed that an external magnetic
field can induce in these systems several features such as metal-insulator transition
\cite{Lu},  Aharonov-Bohm oscillation of conductance \cite{Bach} etc. Inclusion of the Zeeman interaction results in a step-like structure
of the magnetoconductance \cite{Lin} in graphite systems. On the other hand,
electronic properties of the graphite ribbons in the presence of the on-site
Coulomb repulsion are also studied using the Hubbard model within the 
Hartree-Fock Approximation \cite{Fujita, Waka2, Yama}; the Hubbard interaction
is found to favor the formation of ferrimagnetic spin polarization along the
edge of the zigzag
ribbons. Vacancy induced magnetism in graphene ribbons has also been studied 
by using the Hubbard model \cite{Pal}. Recently, the density 
functional theory has been used to determine the electronic and magnetic 
structure of hydrogen-terminated graphene nano-ribbon edges \cite{Wass} 
as well as to 
study the energy gaps and magnetism in {\it bilayer} graphene nano-ribbons in 
the presence of an external electric field between the layers \cite{Sahu}.
Graphenes in the presence of both the 
magnetic and the electric fields have 
also been studied within a noninteracting tight-binding picture by using 
the Green function formalism \cite{Ritter}.     
Few works are also done 
studying the properties of two and three
dimensional graphite lattice structures in the presence of a
magnetic field and the on-site Coulomb repulsion
\cite{Herbut1, Herbut2, Peres}. However, a detailed study of the properties of
the graphite ribbons with different types of edges, in the presence of both
the  magnetic field and the Hubbard interaction is yet to be worked out.\\

In this paper, we investigate how the properties of the graphite ribbons with
zigzag and armchair edges vary with the strength of the magnetic field (both
parallel and perpendicular to the graphite plane) and the on-site Coulomb
repulsion. To understand the role of the special geometry of graphite
structures, we also compare our results with similar calculations on the
square lattice ribbons. We model the system using the Hubbard Hamiltonian in
the presence of a magnetic field, and solve it for the half-filled
band within the unrestricted Hartree-Fock Approximation (HFA).\\

\section{The Model}

For convenience, we use a lattice transformation which changes a hexagonal
lattice to a brick-type lattice, without changing the
lattice topology \cite{Waka1}. Under this transformation the zigzag and the
armchair ribbons look like as shown in Fig.~$1$. We use the following
Hamiltonian to describe our system.

\begin{equation}
 H=\sum_{\langle i,j
 \rangle,\sigma}(t_{i,j}c_{i,\sigma}^{\dag}c_{j,\sigma}+H.c.)+U\sum_{i}n_{i,\uparrow}n_{i,\downarrow}-\vec{B}.\sum_{i,\sigma}\vec{\mu}_{\sigma}n_{i,\sigma},~
\end{equation}

\noindent 
where $c_{i,\sigma}^{\dag}(c_{i,\sigma})$ is the creation(annihilation)
 operator for an electron with spin $\sigma$ at the $i$-th
 site. $n_{i,\sigma}=c_{i,\sigma}^{\dag}c_{i,\sigma}$, and
 $n_{i}=\sum_{\sigma}n_{i,\sigma}$ is the number operator at the $i$-th
 site. $t_{i,j}$ is the hopping integral between the $i$-th and the $j$-th
 sites and $\langle i,j \rangle$ denotes nearest neighbor sites $i$ and
 $j$. Since we are dealing with an ordered lattice, all $t_{i,j}$'s are same
 $(t)$ in the absence of a magnetic field. $U$ is the on-site
 Coulomb repulsion energy. $\vec{\mu}_{\sigma}$ is the moment of an electron
 with spin $\sigma$ and the magnetic moment is hereafter measured in units 
of $\mu_{B}$, the Bohr magneton; $\vec{B}$ is the magnetic field. Under the
 Hartree-Fock Approximation, we decouple the Hubbard term using 
\begin{equation}
U n_{i,\uparrow} n_{i,\downarrow} \rightarrow U \langle n_{i,\uparrow}\rangle
n_{i,\downarrow}+U n_{i,\uparrow} \langle n_{i,\downarrow}\rangle -U \langle
n_{i,\uparrow} \rangle \langle n_{i,\downarrow} \rangle,
\end{equation}

\noindent
where the angular brackets denote the expectation values. This leads to two decoupled
Hamiltonians for the up and the down spins, which can now
be easily diagonalized to obtain the single
particle energy levels in a self-consistent manner. This particular
approximation has been extensively used in studying short ranged electronic
interaction in similar low dimensional systems with reasonable success. Many
of the works on graphite systems treated the electronic correlation using the
unrestricted HFA calculation which yielded physically reliable phases 
\cite{Fujita, Waka2, Yama}. Some novel predictions regarding the edge states of
 the graphenes found in these works are in excellent agreement with the results
 found by calculations based on the Density Functional Theory (DFT) method \cite{Okada, Lee}. 
So the unrestricted HFA has already been well tested
for the present model at least in the limit $\vec{B}=0$. Moreover, the
antiferromagnetic phases driven by the Hubbard correlation in similar low
dimensional systems (e.g. one dimensional C$_{60}$ polymers) has been
successfully studied using the same level of approximation \cite{Hari} to
yield a reasonable comparison with experimental findings. Based on the
above observations it seems meaningful to explore the ground state properties
of our model Hamiltonian $(1)$ using the unrestricted HFA. The stability of
the phases obtained in the numerically self-consistent calculation has been
checked carefully.\\

Throughout our study, we consider periodic boundary condition along the length
of the ribbon and open boundary condition along its width. One can realize
such a boundary condition in practice by wrapping the ribbon in the form of a
cylinder whose axis is parallel to its width. However, we would like to stress
that in our study we have considered the flat ribbon under periodic boundary
condition along its length. As we shall show later on, the results become
insensitive to the boundary condition for system length considered in our
work. We have studied the ribbons of different lengths $(50-150)$
and widths $(3-8)$. For the sake of comparison we have presented the results
for the systems with the same length $(150)$; the widths are varied
as per requirement. We consider two separate
cases, $(i)$ magnetic field parallel to the ribbon plane, and $(ii)$ magnetic
field perpendicular to the ribbon plane. When the magnetic field
$\vec{B}$ is parallel to the ribbon plane, all nearest neighbor hopping
terms along the length of the ribbon will be modified by the same Peierl's
phase $\phi$, such that
\begin{eqnarray}
&&t_{i,j}\rightarrow t e^{2 \pi i \phi/\phi_0},\nonumber\\
&&\phi=\frac{1}{N}\int \vec{B}.\vec{dS}~,
\end{eqnarray}

\noindent
where the surface integral is carried over the cross section of the cylinder. 
$N$ is the length of the ribbon in units of lattice constant. In the
second case, when the magnetic field $\vec{B}$ is perpendicular to the
ribbon plane, we choose the Landau gauge $\vec{A}=(0,Bx,0)$, where we have
assumed the $Y$ axis to be along the translationally invariant direction and
the $X$ axis perpendicular to the $Y$ axis in the ribbon plane. Now the nearest
neighbor hopping integrals along the length of the ribbon will depend on the
$X$ coordinate or the layer number in the following way,   
\begin{equation}
t_{i,j} \rightarrow t e^{2 \pi i \phi_{i,j}/\phi_0}~~,~~~\phi_{i,j}=\int_{i}^{j} \vec{A}.\vec{dl}~~, 
\end{equation}

\noindent
which lead to
\begin{equation}
\phi_{i,j}=\frac{\phi}{2}m~~~~~~~~~~~~\mathrm{for~~ graphite~~ ribbon}, 
\end{equation}

\noindent
and
\begin{equation}
\phi_{i,j}=\phi~ m~~~~~~\mathrm{for~~ square~~ lattice~~ ribbon}, 
\end{equation}

\noindent
where $\phi$ is the flux through a plaquette and $\phi_0$ is the flux quantum; $m$
denotes the layer number of the ribbon containing the $i$-th and the $j$-th
sites.  While performing the transformation of the lattice one ensures that 
the 
flux through one basic hexagonal plaquette in the graphene equals the flux 
through one basic rectangular plaquette in the brick-type lattice. This 
restriction determines the nearest neighbor distance $a$ in the brick-type
lattice in terms of the nearest neighbor distance $b$ in  the graphene ribbon.
We use a typical value for the latter: $b = .25$ nm \cite{Aketa}. In all of 
our calculation we set the scale of energy by setting the nearest neighbor
hopping integral $t=1.0$.\\

For the purpose of illustration we consider the case of $\mu_B B/t = 1.$ For
the graphite ribbons $t$ is $2.7$ev \cite{Oleg}. Value
of the magnetic field $B$ in this situation is $4.6 \times 10^4$ Tesla.\\

We calculate the density of states $\rho(E)=\frac{1}{N}\sum_i \delta(E-E_i)$,
where $E_i$'s are the energy eigenvalues. To study the conductivity of the
system, we calculate the Drude weight $(D)$ and the charge gap $(\Delta)$. The
charge gap $(\Delta)$ at the Fermi level of a system containing $n$ electrons
is given by
\begin{equation}
\Delta = E_{n+1}+ E_{n-1}-2E_{n},
\end{equation}

\noindent
where $E_{n}$ is the ground state energy for a system of $n$ electrons. To
calculate the Drude weight $(D)$, a vanishingly small magnetic flux $\phi'$ is
introduced along the axis of the cylinder-shaped ribbon. The flux $\phi'$ does
not penetrate the ribbon, so it does not alter the Zeeman interaction term and
only modifies the hopping integrals parallel to the length of the ribbon
according to Eq.~$3$. The Drude weight is calculated from the formula 
\cite{Kohn, Scalapino}
\begin{equation}
D=\frac{N}{4\pi^{2}}\left(\frac{\partial^{2}E(\phi')}{\partial\phi'^{2}}\right)_{\phi'=0}~,
\end{equation}

\noindent
where $E(\phi')$ is the ground state energy of the system in the presence of
the flux 
$\phi'$.\\

\section{Magnetic field parallel to the graphite plane}

It can be shown analytically that in the absence of the Hubbard interaction and
the magnetic field, the  armchair ribbons with widths $3M$ and $3M+1$ are
insulating and ribbons with width $3M-1$ are conducting ($M$ is any
integer) at half-filling \cite{Waka1, Nakamura}, we shall present here one case 
of width $3M-1$ (width $5$: $M=2$) and one
case of width $3M$ (width $3$: $M=1$). Cases with large values of $M$ have been 
investigated and are found to yield results which are similar to the ones that 
have been presented here. In Fig.~$2$, we show the density of states
for the armchair ribbons with widths $3$ and $5$ for $U=0$ and $U>0$ in
the absence of the magnetic field. Since at $B=0$, the up and the down spin
bands are degenerate, these diagrams represent the density of states for both
the bands. It is clear from the diagrams that at $U=0$, there is a gap at the
Fermi level for the armchair ribbon with width $3$ at half-filling; this
indicates an insulating behavior, while there is no such gap for width $5$
indicating a metallic or conducting behavior. These show that our numerical
results are in agreement with the analytic results for the armchair ribbons in
the absence of the magnetic field and the Hubbard correlation.\\

For $U>0$, we observe a few
gaps in the energy spectrum other than the Hubbard gap. These gaps are
seen for small ranges of values of $U$, listed below :
\begin{eqnarray}
1.8 \le &U& \le 3.6 ~~~~\mathrm{for~ width}~3,\nonumber\\
3.3 \le &U& \le 4.1 ~~~~\mathrm{for~ width}~5.~~~~~~
\end{eqnarray}

\noindent
We have presented the cases with $U=3$ and $U=3.6$ for armchair ribbons with
widths $3$ and $5$ respectively. These gaps arise due to the special geometry
of the graphite ribbons. It was observed that the position and the size of
these gaps depend on the width of the systems and the magnitude of the 
Hubbard interaction.\\

 Now as we turn on the magnetic field parallel to the
graphite plane, spin reversal symmetry breaks down. The up and the down spin bands
are now non-degenerate. The number of up spins increases in the system at the
cost of the down spins. In the diagram of the density of states, the only
change is the shift of the whole spectrum in the energy scale; it moves
towards left for the up spin bands and towards right for the down spin
bands. The amount of shift depends only on the magnitude of the applied
magnetic field. So it is clear from the diagrams that for certain values of the
magnetic field, the Fermi level will lie within one of the gaps resulting in 
insulating phases and for other values of it the system will be conducting. For
high enough values of the magnetic field the system consists of up spins only
and it becomes spin polarized.\\

In the absence of the magnetic field and the Hubbard correlation, analytic
calculations show that the zigzag ribbons are metallic at half-filling
\cite{Waka1, Nakamura}. In Fig.~$3$, we present the density of states for
zigzag ribbon with width $3$ both in the absence and in the presence of the Hubbard
interaction. It shows that the energy spectrum is gapless for $U=0$
(Fig.~$3(a)$) and for $U>0$ there is only one gap at the Fermi level 
(Fig.~$3(b)$) which is nothing but the Hubbard gap.\\

The Drude weight $(D)$ and the charge gap $(\Delta)$ give a clear idea of the
conductivity of the system. Drude weight becomes high in the
conducting region and zero in the insulating region, while the charge gap is
nonzero in the insulating region and becomes zero or of the order of the
inverse of the system size in the conducting region. Variations of the Drude
weight and the charge gap with the applied magnetic field for a few fixed
values of the Hubbard interaction parameter are shown in Fig.~$4$. First two
diagrams represent the cases for the armchair ribbons with widths $3$ and
$5$ respectively and the third one shows the case for the zigzag ribbon with
width $3$. Though we have presented the variations for a few values of the
Hubbard interaction parameter, similar kind of behavior is observed also for 
other values of it.\\

Using the data for the density of states, the Drude weight and
the charge gap, we have constructed the phase diagrams for these systems in
the $U-B$ plane, which give a detailed idea of the conduction property of
the systems. In Fig.~$5(a)$ and $5(b)$, we have drawn the $U-B$ phase diagrams
of the armchair ribbons with widths $3$ and $5$ respectively. It is clear from
these phase diagrams that for $U=0$, the system is
insulating for width $3$ and conducting for width $5$ in absence of magnetic
field. As the magnetic field is turned on and increased, after a critical
value of $B$, the armchair ribbon with width $3$ becomes conducting. On the 
other hand the ribbon with width $5$ 
continues to be conducting in the presence of $B$. In absence of the magnetic
field the system is insulating due to the Hubbard gap in the spectrum for
$U>0$.  A finite value of the magnetic field is required to turn the system
conducting. At moderate values of $U$ (the range specified in Eq.~$9$),
with the increase of $B$, two additional phase transitions are observed for
both the cases, one from a metallic to an insulating phase and another from an
insulating to a metallic phase. As a result of this, small insulating portions 
are observed in the
middle of the conducting regions of the phase diagrams. These
insulating islands arise due to the gaps other than the Hubbard gaps shown in
the plot of the density of states (Fig.~$2$). These insulating lobes decrease
in size with the increasing
width. It is to be noted here that an increase in the system length (even by an order of magnitude)
does not change the positions and the sizes of the lobes. In fact the results
that we have presented here for system length $N=150$ have already become
independent of the length of the system, and thus practically independent 
of the boundary condition imposed on the system.
At a large value of $B$, the up spin band of the system becomes
completely filled and the down spin band completely empty -- resulting in an
insulating situation for both types of armchair ribbons. The critical value of
$B$ for which the system assumes the spin polarized phase becomes smaller and
smaller with the increase of $U$, because an enhancement of Hubbard
correlation increases the Hubbard gap enormously and greatly reduces the
band width. Though we have not shown the diagrams for armchair ribbons with
width $4$ (or $3M+1$), we have studied it and observed that its behavior is
qualitatively similar to that of width $3$.\\

Figure $5(c)$ represents the $U-B$ phase diagram for a zigzag ribbon with width
$3$. For $U=0$ the zigzag ribbon is conducting as there appears no gap at the
Fermi level. Nonzero $U$ opens a gap at the Fermi level making the system
insulating. For a small value of $U$, only a small value of $B$ is sufficient
 to drive the system conducting, while a higher value of $U$
 requires a higher value of $B$ for the transition. As we have
seen in the case of the armchair ribbons, in this case also a high enough value of
$B$ makes the system spin polarized. So for the zigzag ribbons,
application of magnetic field parallel to the graphite plane causes one phase
transition from an insulating to a conducting one and another from a
conducting to an insulating spin polarized one for nonzero values of $U$. In
this case, no other insulating lobe at the middle of the conducting region is
obtained, since there is no gap in the energy spectrum other than the Hubbard 
gap.\\

This is interesting to note that only in the case of the armchair ribbon there 
arise additional gaps apart from the usual Hubbard gap (see, for example, the 
gap around $E=3.5$ in Fig.~$2(b)$). In the presence of the Hubbard correlation 
(i.e. $U\neq0$) the system is driven to a spin density wave (SDW) state. If 
one moves along the edge of the ribbon one should encounter a distribution of 
moments corresponding to a dominant SDW with wave vector $q=\pi$. This leads
to the formation of up and down spin moments at the neighboring sites which
look like a one dimensional antiferromagnetic modulation. It is well known that such an SDW
modulation is responsible for opening up Mott-Hubbard gap at the Fermi level of
a half-filled electronic band in a bipartite lattice. However, in the 
case of an armchair ribbon we find an additional underlying modulation of 
moments corresponding to an SDW with $q\neq\pi$. It can be traced by observing
the repetition of the peaks of $S_i$ of same height in Fig.~$6(a)$. This type
of modulation is not observable in 
zigzag ribbons (Fig.~$6(b)$). The additional SDW modulation is responsible for 
opening up additional gaps in the spectrum of the armchair ribbons in case 
of $U\neq0$. Certainly these gaps will decrease with increasing ribbon width 
but will remain sensibly unchanged with increasing length. As we have noted in
Eq.~$9$ these gaps are observable for a narrow range of value of $U$ that
depends on the width of the system. Larger $U$ values wipe out the effect of
the additional SDW modulation and consequently the usual Mott-Hubbard gap
alone survives.\\  

We check the sensitivity of our results to the system length by calculating a
quantity $\Sigma$ defined by $\Sigma =\left[ \sum_{i=1}^{N}
  \frac{\left(\sigma_{i}^{\rm pbc} -\sigma_{i}^{\rm
      obc}\right)^2}{N}\right]^{\frac{1}{2}}$, where $\sigma_i =
n_{i,\uparrow} - n_{i,\downarrow}$, the moment at site $i$ along the symmetry
axis of the ribbon. The variation of $\Sigma$ with the system length $N$ is
plotted in Fig.~$6(c)$ for a typical value of $U$ $(= 2.4)$ and $B$
$(= 1.0)$. It turns out that the results become insensitive to the boundary
condition as $\Sigma$ goes to zero for $N \sim 100-150$.

\section{Magnetic field perpendicular to the graphite plane}

In this section similar studies are done but the direction of the magnetic
field is taken perpendicular to the ribbon plane. The effect of the magnetic
field $B$ is quite complicated in this case. As we
have discussed earlier (Eqs.~$4$ and $5$), for this type of magnetic
field the hopping integrals in different layers are modified differently. It
causes the formation of multiple bands (depending on the magnitude of the
applied magnetic field) in the energy spectrum. On the other hand, the Zeeman
term present in the Hamiltonian shifts the energy spectrum for the
up and down spin bands in opposite directions. These two processes together
contribute to the properties of the system. In Fig.~$7$, we show the density
of states of the armchair ribbons with widths $3$ and $5$ for $U=0$ in the 
presence of the magnetic field $B$. For
both $U=0$ and $B=0$, we have seen the nature of the density of states
earlier in Fig.~$2$. Here we see that for nonzero $B$, gaps open at
different parts of the energy spectrum. We have shown the case with
$B=1.65$. Even for $U>0$ this type of gaps are observed. In Fig.~$8$,
the density of states for finite $U$ is
shown for the same value of the magnetic field.\\

Next we present the cases of the zigzag ribbons with width $3$. Figures~$9(a)$
and $9(b)$ show the nature of the density of states for $U=0$ and $B=1.65$. In
this case also multiple gaps appear in the energy spectrum. Figures~$9(c)$ 
and $9(d)$ show the same for $U=2$.\\

With the magnetic field perpendicular to the ribbon
plane, the positions of the gaps in the energy spectrum depend
on the magnitude of the magnetic field, unlike the case with the magnetic
field parallel to the ribbon plane. Studying the variations of the Drude
weight and the charge gap with $B$ for fixed $U$ values, we see that a few
transitions occur between the insulating and the conducting phases with the
increase of $B$ from the zero value. Figure~$10$ shows one representative
case (for armchair ribbon of width $3$ at $U=1$) of such variations.\\

To get a
clear view, we have plotted the $U-B$ phase diagrams of the systems in
Fig.~$11$. Here the first two diagrams are for the armchair
ribbons with widths $3$ and $5$ respectively and the third one is the same for
the zigzag ribbon of width $3$. For the purpose of comparison we have also
presented the phase diagram for the square lattice ribbon in the same
panel. It shows that for a fixed value of $U$ (weak to intermediate coupling)
there appear small domains along the vertical direction (width depending on
the value of $U$) where the system becomes metallic. Thus we find some 
conducting lobes immersed in
the ``sea'' of a large insulating region. These metallic lobes are marked with 
even Roman numerals (e.g. II, IV and VI) in Fig. 11. Apart from these there 
also appear tiny metallic lobes in the case of the armchair ribbon of width
$5$ (within the insulating region V in Fig. 11(b)). In the case of the zigzag
ribbon, on the contrary, we find a very small insulating lobe inside a
conducting one (i.e. lobe number IV in Fig. 11(c)).  Such tiny lobes are
observed to be finite size effects and generally disappear with increasing
length/ width of the system. It is to be noted here that within these metallic
regions there appears a partial spin polarization that increases with the
increase in the magnetic field strength. Due to the existence of multiple 
metallic lobes
one can observe, in principle, several Metal-Insulator (MI) 
transitions driven by the external magnetic field.  Observing first one
or two of such transitions may be experimentally feasible in the case of the 
armchair configuration, although, in general, the value of $B$ required to see 
all these transitions will be too high compared to the present experimental 
facility. It is seen from Figs. 11(a) and 11(b) that in the case of armchair
ribbons, for large enough  $U$, the width of the conducting lobe (region II)
is so small that in this region two consecutive MI transitions may take place
with a small increase in $B$, a feature that may be interesting in view of
some 
field-driven switching mechanisms.\\

On the other hand the insulating region shows a rich variety of phase structure
controlled by the values of the parameters as well as crucially controlled by
the boundary effects and system widths. Generally,  after a large enough value
of $B$ the system becomes spin polarized (e.g. region V in Figs. 11(a)-(c))  beyond which no further phase transitions
occur with increasing $B$. The first insulating region (region I in
Figs. 11(a)$-$(c)) obtained for small  values of $B$ shows some magnetic
structure for $U\neq 0$. For the armchair ribbons this region shows an
antiferromagnetic modulation of electron spins in a given layer. However, this
antiferromagnetic phase is not homogeneous as usually observed due to a Hubbard
interaction on an infinite bipartite lattice. Here, by moving from one layer
to an adjacent layer, one finds that the sublattice magnetization per  layer
strongly depends on the layer index (controlled essentially by the proximity
of the boundary). For a ribbon with odd number of layers (3, 5 or 7) there
always appears an ``inversion'' symmetry of the patterns about the central
layer. For ribbons of even number of layers too such a symmetry exists about
the central horizontal axis (passing through midway between two middle layers
of atoms). In case of a zigzag ribbon, however, this insulating region shows
an antiferrimagnetic alignment along the layers at or near the boundary, and a
perfect antiferromagnetic alignment only at the central layer (for odd
width). The reason behind this discrepancy between the zigzag and the armchair
configuration is much similar to that explained in
the context of Fig. 6. The other insulating patch that is interspersed
between the conducting lobes (e.g. region III) have
partial spin polarization forming a net  magnetization in the system (due
to an appreciably large value of $B$). In the case of armchair ribbons one
finds a spin distribution much like a ferro-arrangement, magnetization in a
layer decreasing from the boundary to the central layer (obeying the previously
mentioned inversion symmetry). In the case of the zigzag ribbon the moment
distribution pattern is ``ferrimagnetic'' in a layer. Again the variation  of
the moment distribution from the boundary layer to the central one shows the
aforesaid ``inversion'' symmetry. This shows that in the presence of a
magnetic field the same hexagonal network manifests the effect of Coulomb
correlation in  different ways subject to the presence of boundaries of
different geometries at finite distances. 

In the special case of an armchair ribbon of width 3 the insulating region I
persists even upto $U=0$ (known to be an exact result at least for $B=0$
\cite{Waka1, Nakamura}) where the system behaves like a paramagnetic
insulator. Antiferromagnetism is switched on as soon as $U$ becomes
non-zero. For the armchair ribbon of width 5, however, the system is
conducting at $U=0$, $B=0$. Antiferromagnetism sets in for $U>0$ in the
absence of the magnetic field. In this
later case the antiferromagnetic order parameter shows a sudden increase in
its value around $U/t \sim 2.2$ for $B=0$. However, within the present mean
field calculations  one cannot predict a critical value of $U/t$, even if it
is there, because the  charge gap and the antiferromagnetic order parameter
truly vanish only at $U=0$.  

\section{Comparison between the graphite and the square lattice ribbons}

To understand the effect of the special type of structures of graphite 
lattices, we have
 studied for comparison the square lattice ribbon structures of width $3$ at
half-filling. These systems are metallic in the 
absence of the Hubbard interaction and the magnetic field. Non-zero
 Hubbard interaction makes the system insulating. It is clear from the
plot of the density of states with $B=0$ (Fig.~$12$)
that the $U-B$ phase diagram (when $B$ is parallel to the ribbon plane) will
be qualitatively similar to that of the  zigzag graphene ribbons. 
In Fig.~$13$, the density of states of the square lattice ribbons of width $3$
in the  presence of the magnetic field perpendicular to the ribbon plane is
presented. Here also multiple gaps open up in the energy
spectrum. Fig. $11(d)$ shows the $U-B$ phase diagram for this case. From
Fig. $11$  it is revealed that the phase diagram of the square ribbon resembles
closely that of the armchair graphene rather than the that of the zigzag
ribbon. Also 
the nature of spin polarization in the insulating regions for the square
lattice is found to be much similar to the case of the armchair ribbons. In
view of this above mentioned similarity between the armchair and square
lattice ribbons we further note that the sites on the boundary layers of each
of these ribbons belong to the similar status with respect to the coordination
numbers. On the other hand the sites of the boundary layer of a zigzag ribbon
are not at the same status (see Fig.1).\\

\section{Conclusion}
We have studied the simultaneous effects of the magnetic field and the on-site Coulomb
interaction on the nanographite ribbon structures. Single orbital
nearest neighbor tight-binding
model has been used to study the systems. Electronic correlation is included in
the Hamiltonian using the Hubbard interaction. Reasonably large systems are
studied in this work. We have used 
the unrestricted Hartree-Fock Approximation method, which was found to be
reliable in these type of systems \cite{Fujita, Waka2, Yama}. Since a magnetic
field in any direction can be resolved into two components, one parallel and
another perpendicular to the graphite plane, we have studied the effects of the
magnetic fields in the two above mentioned directions independently. In
nanographite structures, shapes of the boundary play an important role in
determining 
the behavior of the systems. Here we have considered two important boundary
geometries - the zigzag and the armchair, and studied the effect of the
magnetic field on the ribbon structures under periodic boundary condition
along their lengths. We have calculated the density of states for different $U$
and $B$ values. Studying these density of states, the charge gap and the Drude
weight we obtained the $U-B$ phase diagrams for different types of graphite
ribbons. To get a clear understanding of the geometry effect, we have
compared the results of the graphite ribbons with that of the square lattice
ribbons. With the magnetic field parallel to the ribbon plane these phase
diagrams show that the behavior of the armchair ribbons is markedly different
from those of the zigzag and the square lattice ribbons. For armchair ribbons
small insulating islands at the middle of the conducting regions of the phase
diagrams are obtained which are absent in other cases. The reason behind this 
difference in behavior of armchair and zigzag ribbons is also qualitatively 
explained.  A close comparison between the case of armchair and zigzag ribbons
in the presence of a magnetic field reveals that the same hexagonal graphene
network manifests the effect of the Hubbard interaction in different ways
controlled by the boundary topology. When the magnetic field
is perpendicular to the ribbon plane, the phase diagram contains several
conducting  lobes submerged in an insulating sea. Therefore, such systems
would show multiple metal-insulator transitions tuned by the magnetic field
although  much higher values of the field would be required to see these
transitions than  are commonly realizable in the laboratories at present.
 In view of such multiple metal-insulator transitions the 
graphite ribbons may find their practical application in designing 
field-driven switching devices. Further study of the graphite ribbons with 
other boundary
geometries may yield some interesting results. Also the effect of the finite
temperature on the graphite ribbons for the competing regime of the correlation
parameter and the magnetic field may be interesting.\\  

\noindent
{\bf acknowledgments:}\\
One of the authors (BB) was supported by the Jadavpur University Research
Grant from the Jadavpur University, Kolkata.

\newpage
{\Large{\bf {Figure captions}}}

Fig 1: The transformation from hexagonal to brick-type lattice, without
changing the lattice topology.\\

Fig 2: Plot of the density of states (D.O.S) of the armchair ribbons of 
length $150$ in absence of magnetic field. The widths (w) of the ribbons and
the magnitude of the Hubbard interaction $(U)$ are specified on the diagrams.\\

Fig 3: Plot of the density of states of the zigzag ribbons of length $150$ in
  absence of magnetic field. The widths (w) of the ribbons and the magnitude
  of the Hubbard interaction $(U)$ are specified on the diagrams.\\

Fig 4: Plot of the Drude weight $(D)$ and the charge gap $(\Delta)$
  as functions of magnetic field (parallel to ribbon plane) for the armchair
  and the zigzag graphite ribbons (length $150$) at half-filling; dotted line
  is for the Drude weight and solid line is for the charge gap. The type and
  the widths (w) of the ribbons and also the magnitude of the Hubbard
  interaction $(U)$ are specified on the diagrams.\\

Fig 5: $U-B$ phase diagrams of the armchair and the zigzag ribbons of length
  $150$ when the magnetic field is parallel to the graphite plane at
  half-filling. The widths (w) and the type of the ribbons are mentioned on
  the diagrams. The ``ins'' in the second diagram means insulating region.\\

Fig 6: Plot of $S_i=n_{i\uparrow}-n_{i\downarrow}$, the moment at site $i$ 
along the upper edge, as a function of the site index $i$ for graphite ribbons 
with (a) armchair boundary and (b) zigzag boundary at $U=3.$ and $B=0.$. For
simplicity we consider ribbons with width $3$ and length $20$. Features are
similar for larger systems. Results given in (b) closely resemble those found
using DFT calculations \cite{Okada, Lee}. In $(c)$ we have plotted $\Sigma$ as
a function of system length $N$. This graph checks the insensitivity of our
result to the boundary condition for our system length $150$.\\

Fig 7: Plot of the density of states of the armchair ribbons of length $150$
  in presence of magnetic field $(\mu_B B/t=1.65)$ perpendicular to graphite
  plane when  $U/t=0$. Widths (w) of the ribbons are specified on the
  diagrams.\\

Fig 8: Plot of the density of states of the armchair ribbons of length $150$
 in presence of magnetic field $(\mu_B B/t=1.65)$ perpendicular to graphite
 plane and Hubbard interaction $(U/t=2)$. Widths (w) of the ribbons are
 specified on the  diagrams.\\

Fig 9: Plot of the density of states of the zigzag ribbons of length $150$
  and width $3$ in presence of magnetic field $(\mu_B B/t=1.65)$ perpendicular
  to the graphite plane. On-site Coulomb correlation parameter values are
  mentioned on the diagrams.\\

Fig 10: Plot of the Drude weight $(D)$ and the charge gap $(\Delta)$
  as functions of magnetic field (perpendicular to ribbon plane) for the
  armchair graphite ribbon of length $150$ and width $3$ at half-filling;
  dashed line is for the Drude weight and solid line is for the charge
  gap. The magnitude of the Hubbard interaction $(U)$ is $1$.\\

Fig 11: $U-B$ phase diagrams of armchair ((a) and (b) for widths $3$ and $5$
  respectively), zigzag ((c) for width $3$) and square lattice ribbons ((d)
  for width $3$) of length $150$ when the magnetic field is perpendicular to
  the ribbon plane, at half-filling.\\

Fig12: Plot of the density of states of the square lattice ribbons of width
  $3$ and length $150$ in absence of magnetic field. On-site Coulomb
  correlation parameter values are mentioned on the diagrams.\\

Fig 13: Plot of the density of states of the square lattice ribbons of length
  $150$ and width $3$ in presence of magnetic field $(B=1.65)$ perpendicular
  to ribbon plane. On-site Coulomb correlation parameter values are mentioned
  on the diagrams.\\

\newpage
\Large{\bf {Figures}}

\begin{figure}[h]
\resizebox{6.5cm}{!}
{\includegraphics*{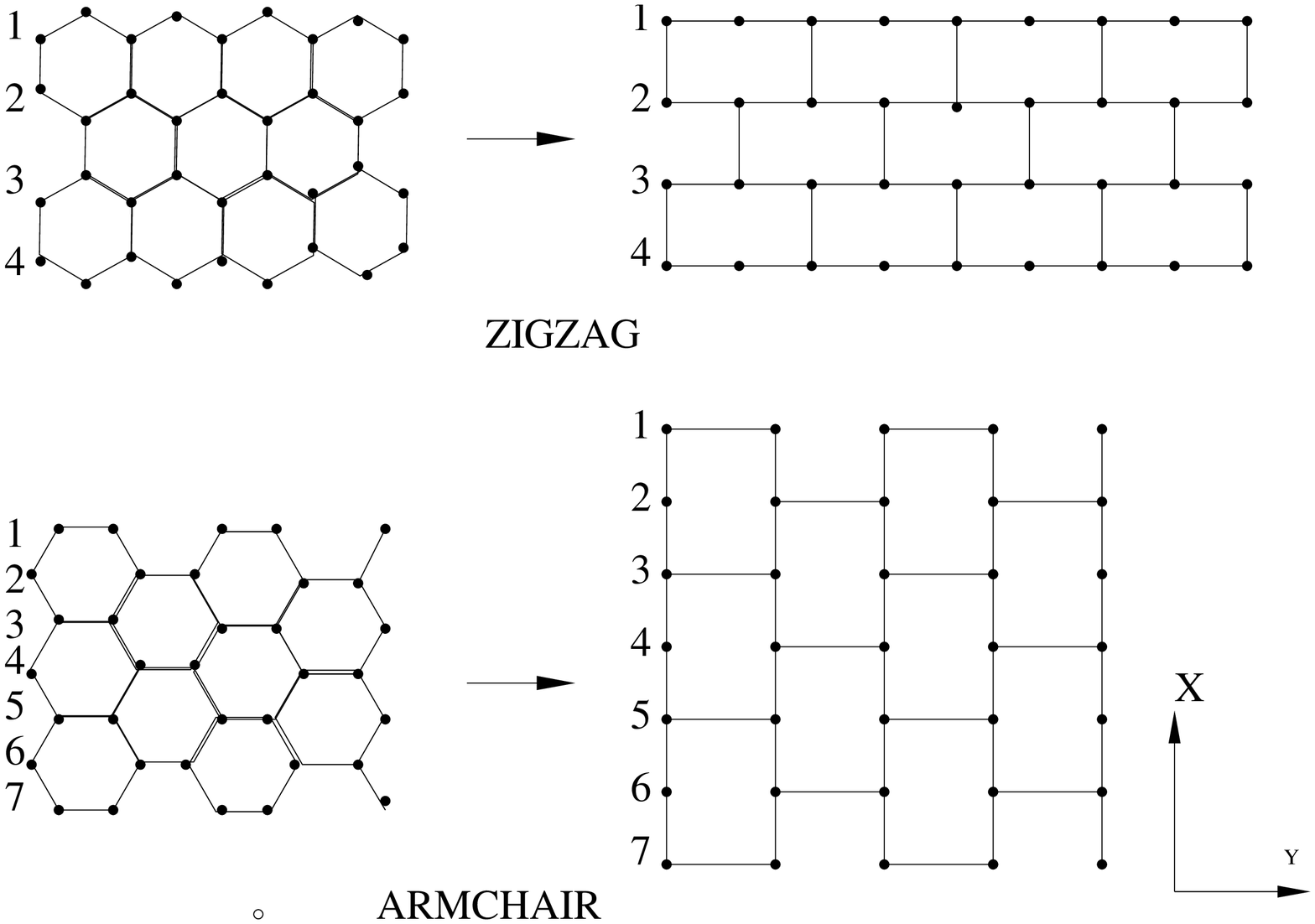}}
\caption{}
\end{figure}

\newpage
\begin{figure}[h]
\resizebox{6.5cm}{!}
{\includegraphics*{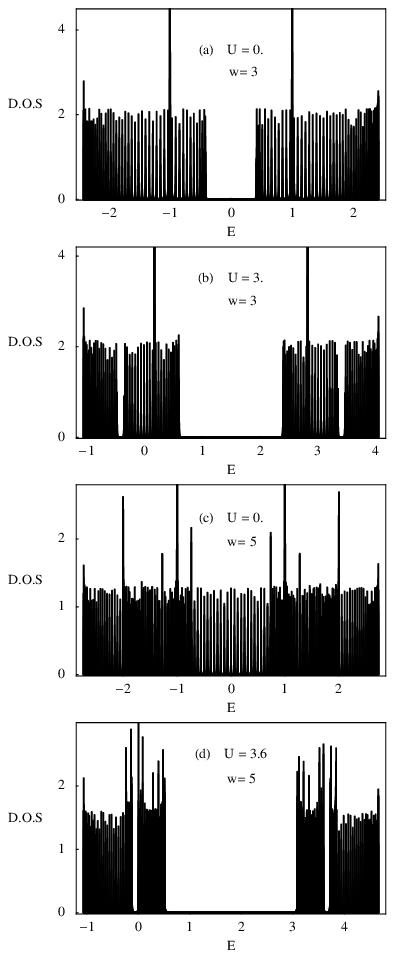}}
\caption{}
\end{figure}

\newpage
\begin{figure}[h]
\resizebox{6.5cm}{!}
{\includegraphics*{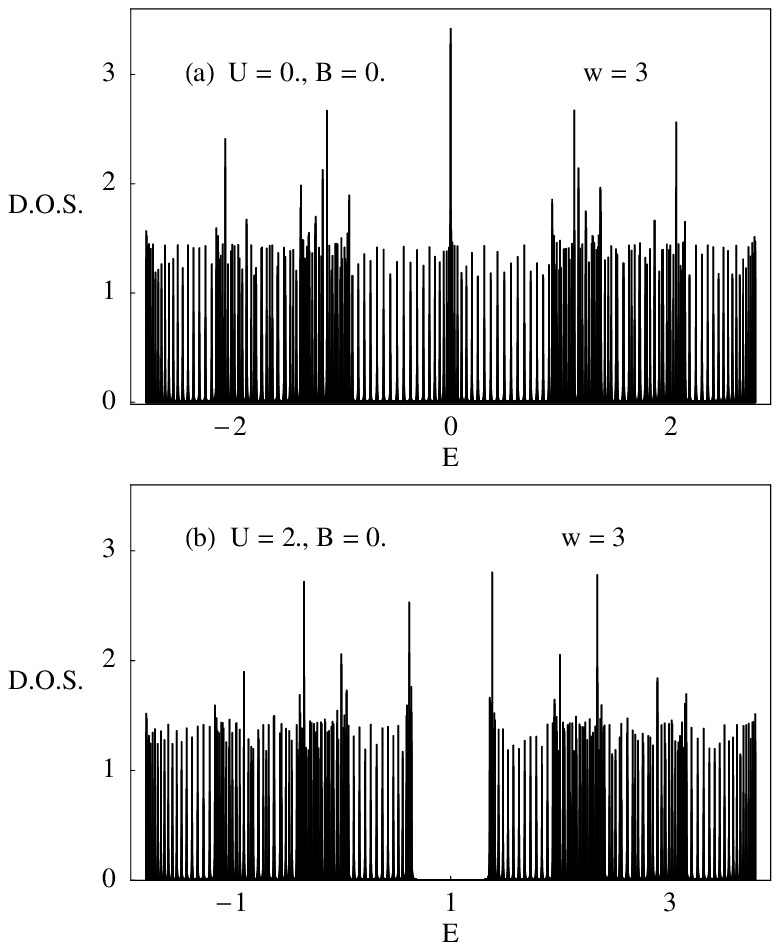}}
\caption{}
\end{figure}

\newpage
\begin{figure}[h]
\resizebox{6.5cm}{!}
{\includegraphics*{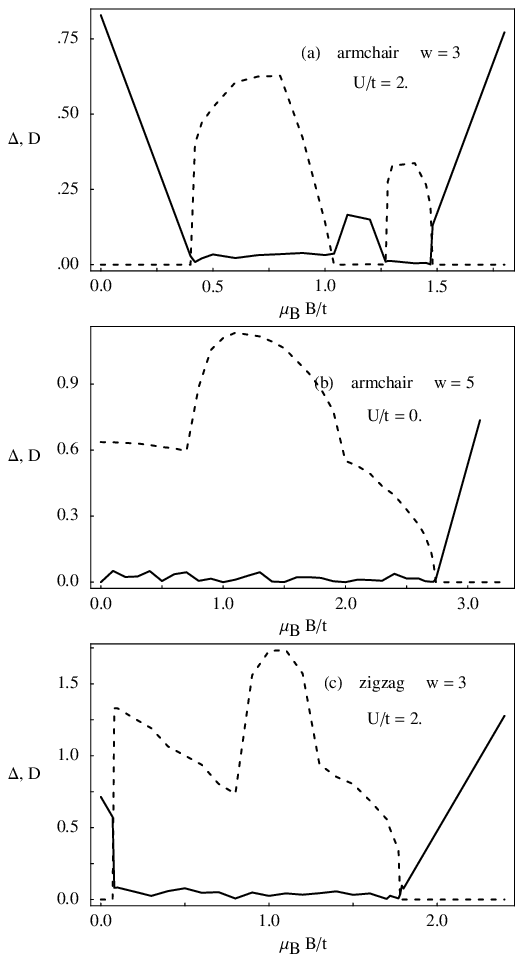}}
\caption{}
\end{figure}

\newpage
\begin{figure}[h]
\resizebox{6.5cm}{!}
{\includegraphics*{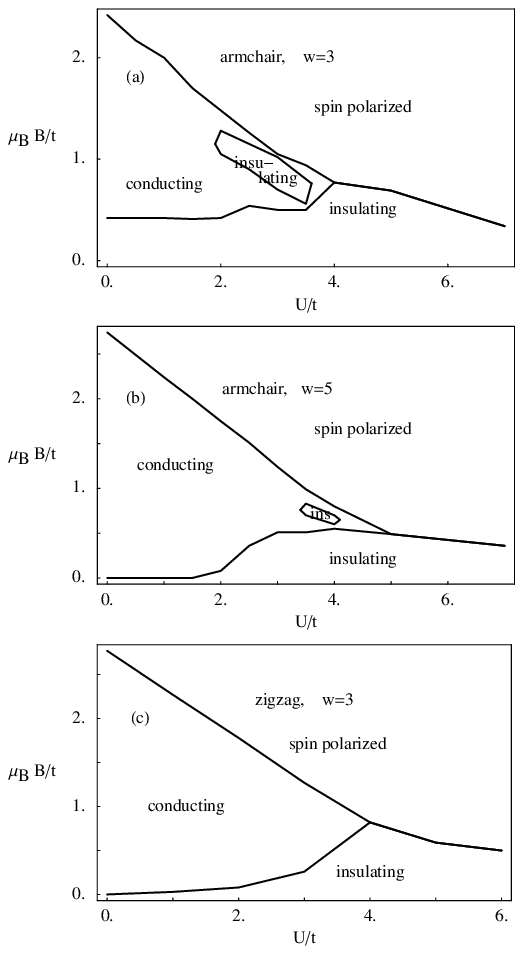}}
\caption{}
\end{figure}

\newpage
\begin{figure}[h]
\resizebox{6.5cm}{!}
{\includegraphics*{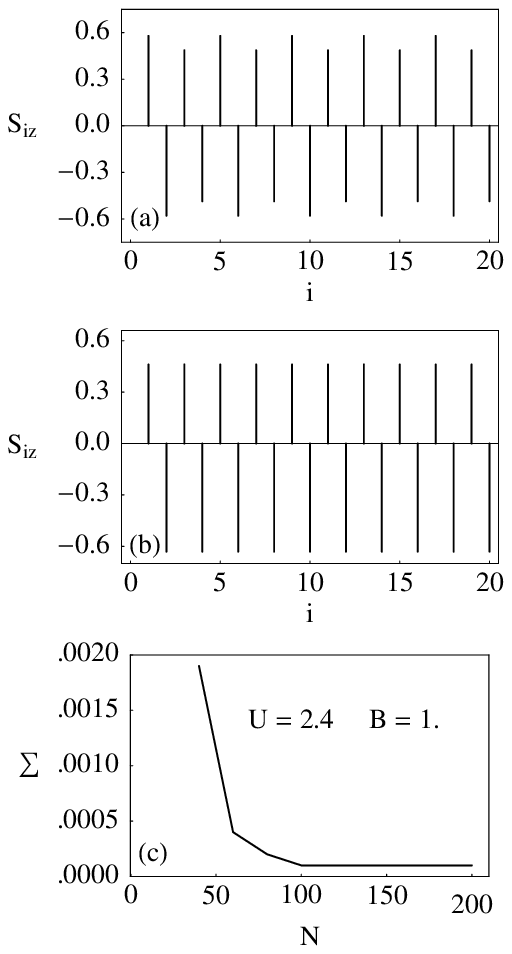}}
\caption{}
\end{figure}

\newpage
\begin{figure}[h]
\resizebox{6.5cm}{!}
{\includegraphics*{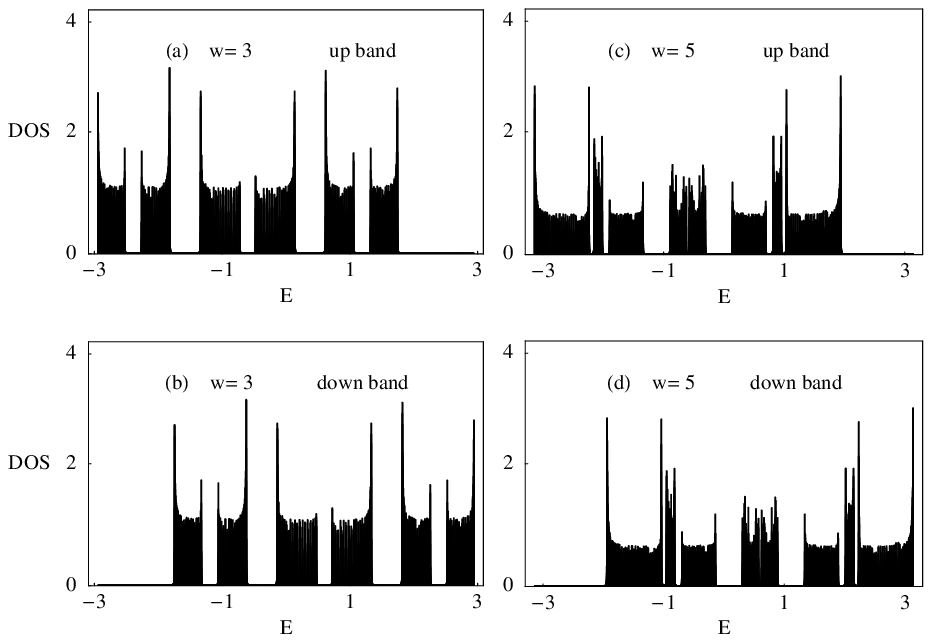}}
\caption{}
\end{figure}

\newpage
\begin{figure}[h]
\resizebox{6.5cm}{!}
{\includegraphics*{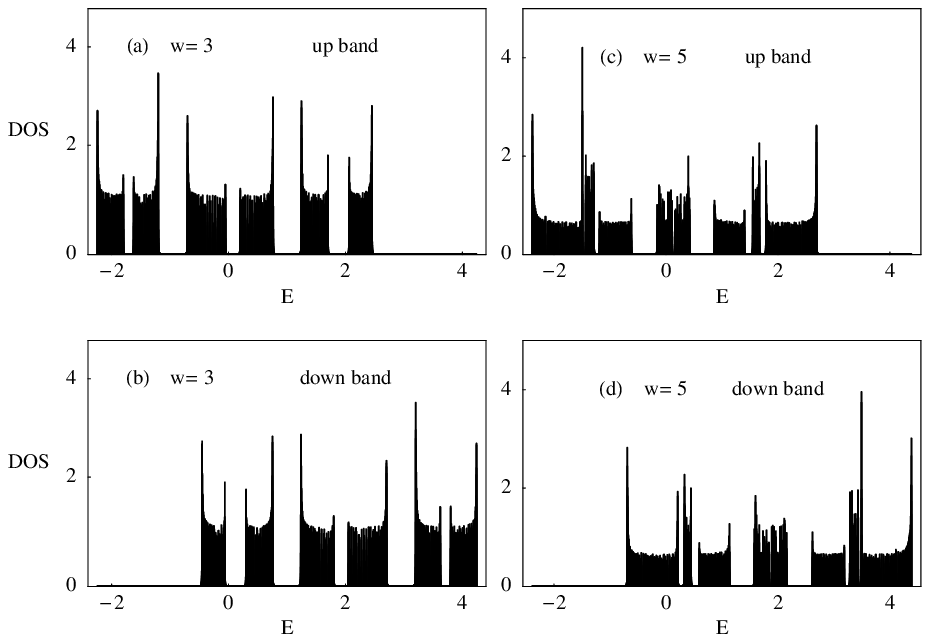}}
\caption{}
\end{figure}

\newpage
\begin{figure}[h]
\resizebox{6.5cm}{!}
{\includegraphics*{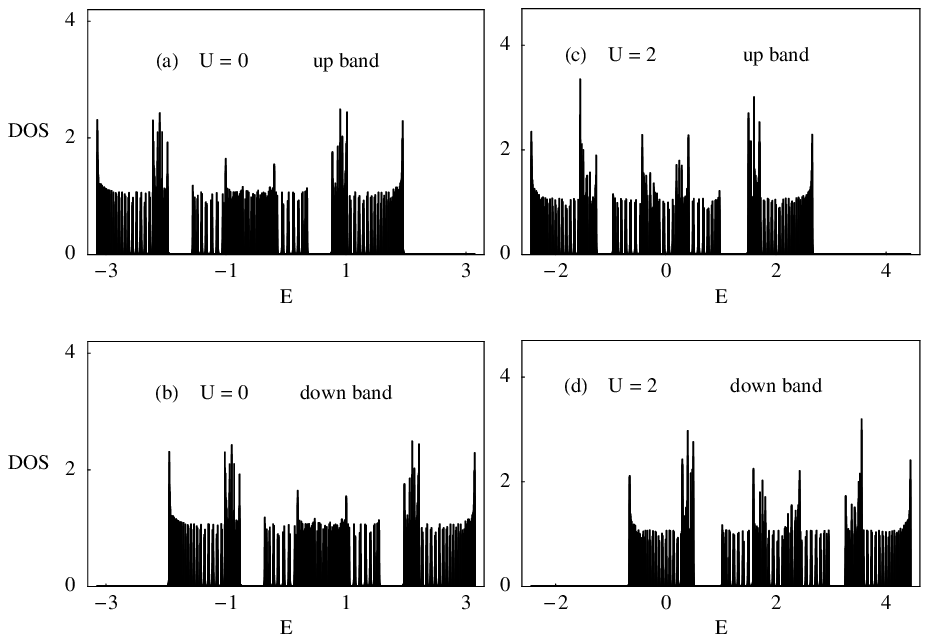}}
\caption{}
\end{figure}

\newpage
\begin{figure}[h]
\resizebox{8cm}{!}
{\includegraphics*{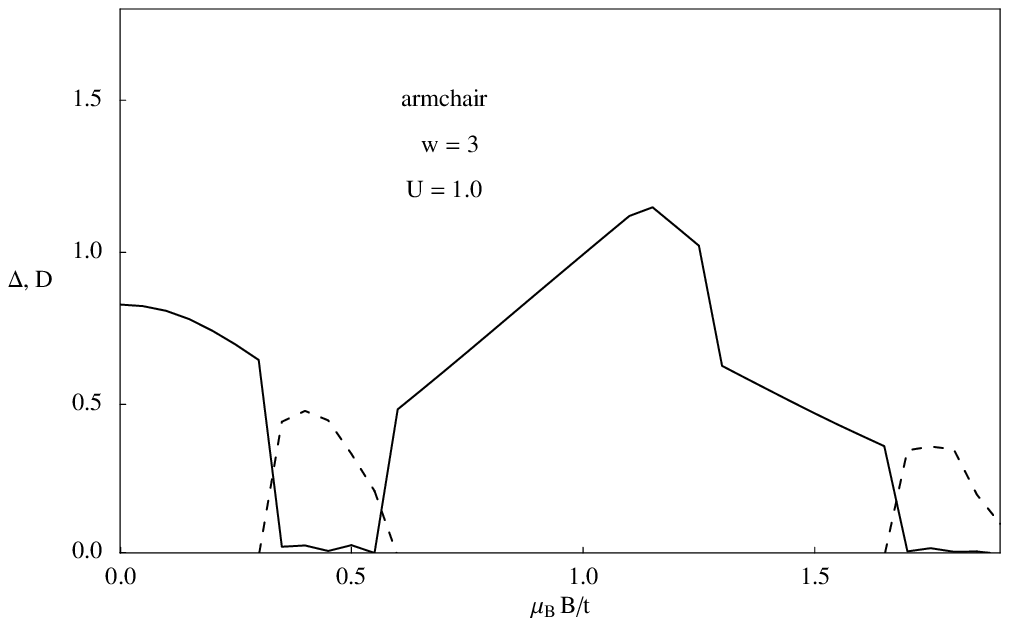}}
\caption{}
\end{figure}

\newpage
\begin{figure}[h]
\resizebox{8cm}{!}
{\includegraphics*{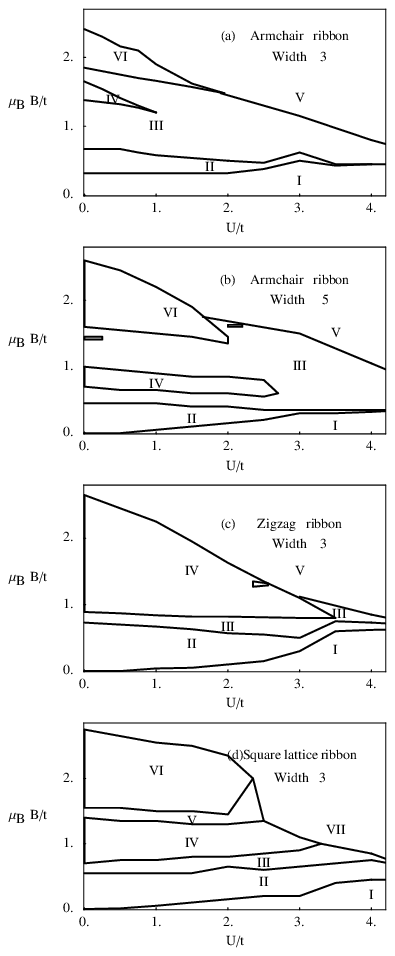}}
\caption{}
\end{figure}

\newpage
\begin{figure}[h]
\resizebox{6.5cm}{!}
{\includegraphics*{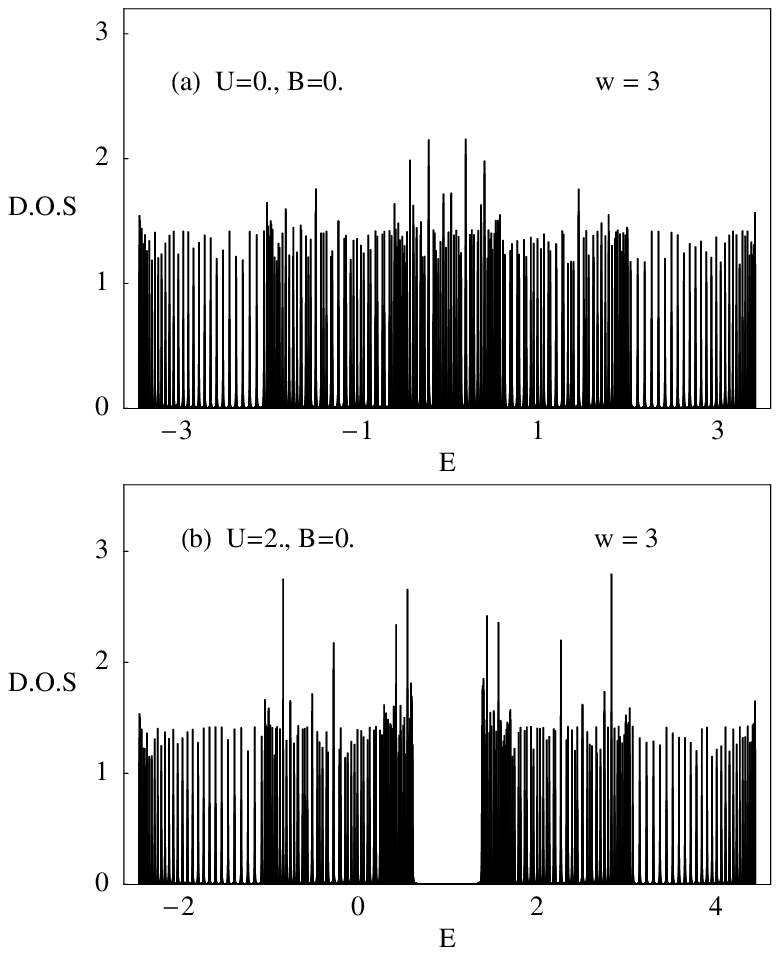}}
\caption{}
\end{figure}

\newpage
\begin{figure}[h]
\resizebox{8cm}{!}
{\includegraphics*{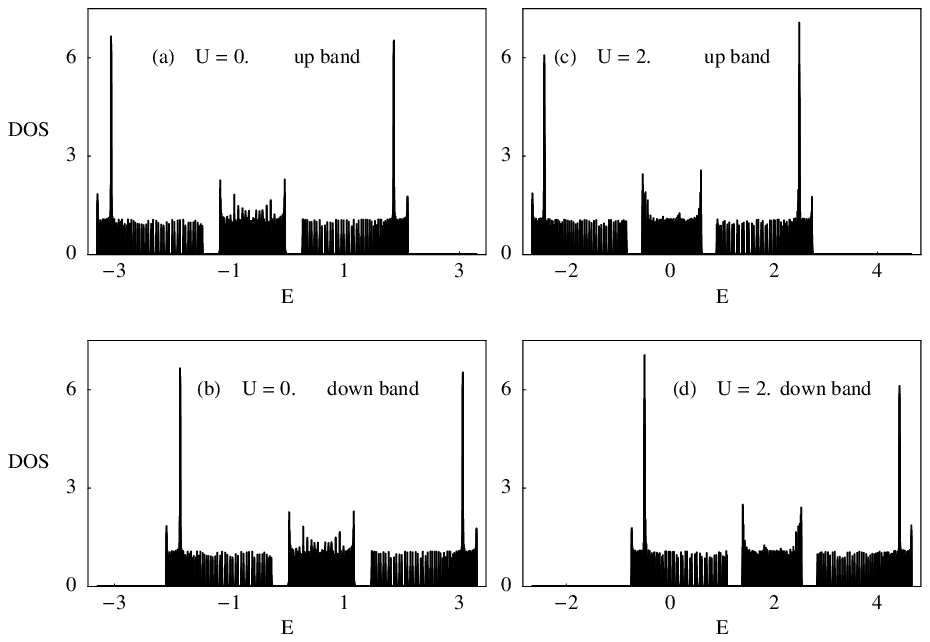}}
\caption{}
\end{figure}

\end{document}